\documentclass[10pt,onecolumn,letter]{IEEEtran}
\usepackage{graphicx,latexsym}
\usepackage{amssymb}
\usepackage{times}

\title{Impact of Interleaver Pruning on Properties of Underlying Permutations}

\author{Fred~Daneshgaran,~\IEEEmembership{Member,~IEEE,} Marina Mondin,~\IEEEmembership{Member,~IEEE,}
\thanks{Manuscript received...; revised...}%
\thanks{F. Daneshgaran is with the ECE Dept.,
Calif. State Univ., Los Angeles, USA.}
\thanks{M. Mondin is with the Dipartimento di Elettronica e Telecomunicazioni, Politecnico di
Torino, Italy.}}

\begin{document}\maketitle

\begin{abstract}
In this paper we address the issue of pruning (i.e., shortening) a given interleaver via 
truncation of the transposition vector of the mother permutation and study its impact
on the structural properties of the permutation. This method of pruning allows for continuous
uninterrupted data flow regardless of the permutation length since the permutation engine is a buffer 
whose leading element is swapped by other elements in the queue.
The principle goal of
pruning is that of construction of variable length
and hence delay interleavers with application to iterative soft information processing and 
concatenated codes, using the same structure (possibly in hardware)
of the interleaver and deinterleaver units. We address the issue
of how pruning impacts the spread of the permutation and also look at how pruning impacts
algebraically constructed permutations. We note that pruning via truncation of the transposition vector
of the permutation can have a catastrophic impact on the permutation spread of algebraically constructed
permutations. To remedy this problem, we propose
a novel lifting method whereby a subset of the points in the permutation map leading to low spread
of the pruned permutation are identified and eliminated. Practical realization of this lifting is then proposed
via dummy symbol insertion in the input queue of the Finite State Permuter (FSP), and subsequent removal of the dummy symbols
at the FSP output.

\end{abstract}


\section{Introduction}

\PARstart{I}{nterleavers} appear in many modern communication link designs at the physical
layer. Their primary function is to scramble and de-correlate the data at the interleaver
output, permitting the generation of extrinsic information needed in many iterative signal processing paradigms
that are by now a standard procedure to achieve performances at the physical layer near the theoretical limits.  
They have a long history but one of their first systematic uses were in connection with concatenated channel codes [1-26]. 
An example of this is in Parallel Concatenated Convolutional Codes (PCCC or
Turbo codes) which are potential candidates for many applications due to
their excellent performance near the capacity limit
\cite{ber,unveil,div1}. Upon proper
termination, the PCCC can essentially be interpreted as a large
block code which we denote as $C_{p}$.

The principle goal of
pruning is that of construction of variable length
and hence delay interleavers with application to iterative soft information processing and 
concatenated codes. In a series of recent papers \cite{mans1, mans2, mans3} M. M. Mansour has developed a systematic and efficient technique for
determination of permutation inliers that can be used for construction of variable length interleavers. Our
approach is fundamentally different. 

In \cite{fred1} we presented a systematic
iterative technique for construction of interleavers tailored to
given constituent Recursive Systematic Convolutional (RSC) codes leading to significant performance
gains for a given length interleaver. This improvement is
manifested in the distance spectrum of $C_{p}$, and the resulting
interleavers lead to PCCCs with better distance spectra compared
to other interleaver design techniques proposed in the literature.

In \cite{fredprune} we focused on the design of variable length
interleavers with application to turbo codes, starting from {\em
any} generic interleaver of maximal length. We presented a pruning
method suitable for practical implementation and applicable to any
interleaver and demonstrated the average optimality of the pruning
strategy. Variable length block coding may be of significant
interest in practice since a major component of the Quality Of
Service (QOS) is the delay. Since the encoding and decoding delay
associated with turbo codes are essentially dominated by the block
length (or the data frame length), it is of considerable practical
interest to be able to construct variable length Turbo codes.

This paper explores the theoretical properties of the general pruning method presented
in \cite{fredprune} and addresses the impact of pruning on permutation spread and on structured algebraically constructed
permutations. Our focus on algebraic constructions will be on Quadratic Permutation Polynomials (QPP) \cite{take1,take2}
and we shall demonstrate that general pruning as proposed in \cite{fredprune} can have catastrophic
impact on permutation spread. We then present a novel technique of lifting permutation points leading
to low spread via dummy symbol insertion in the
sequence to be interleaved to remedy this problem.

There are several important features of the pruning method  presented in \cite{fredprune} that we wish to recall
before we proceed with the rest of the paper:  
\begin{enumerate}
\item the technique presented in \cite{fred1} is an interleaver
growth algorithm (i.e., opposite of pruning) with one major
difference with the pruning strategy presented in \cite{fredprune}. During the
construction of an interleaver of length $K+1$ from one of length
$K$, one has a {\em choice} of $K+1$ interleavers to select from
in order to minimize some appropriate cost function. In pruning an
interleaver of length $K+1$ to one of length $K$ using the method
proposed in \cite{fredprune}, one has only {\em one choice}. Hence,
growing an interleaver via the method of \cite{fred1}, one ends up
with an implicitly prunable optimized (in a narrow sense)
interleaver. Starting with an arbitrary interleaver, the pruning
may not necessarily lead to optimal results in any strict sense.
What we have shown is {\em average optimality} of the approach in the context
of channel coding;
\item the technique presented in \cite{fredprune} and summarized below allows any
interleaver performing a given permutation to be reduced in size
with gradual performance degradation of the resulting PCCC
consistent with what one would obtain via redesign of the
interleaver each time, but without the need for changing almost
completely the operation of the interleaver.
\end{enumerate}

The rest of the paper is organized as follows. In section~II we provide a review
of the transposition realization of any given permutation and its use
in realization of the interleaver as a Finite State Permuter (FSP) originally
introduced in \cite{fred1}. In section~III we present the core results on interleaver pruning and
growth and how this process impacts the underlying permutations. We also address the issue
of how pruning impacts the spread of the permutation. Finally we look at how pruning impacts
algebraically constructed permutations. We note that pruning via truncation of the transposition vector
of the permutation can have a catastrophic impact on the permutation spread. To remedy this problem, we propose
a novel lifting method whereby a subset of the points in the permutation map leading to low spread
of the pruned permutation are identified and eliminated. Practical realization of this lifting is then proposed
via dummy symbol insertion in the input queue of the FSP, and subsequent removal of the dummy symbols
at the FSP output. Section~IV is devoted to conclusions.  

\section{The Finite State Permuter and Transposition Vector of a Permutation}

Design of concatenated signal processing schemes and in particular, Parallel or Serial concatenated codes, 
is tied to the
realization of the interleaver and deinterleaver. One approach
for implementation is that of Random Access Memory (RAM) units whereby the data are
written in a particular order and read according to the permutation to be implemented
by the interleaver, and its inverse to be implemented by the deinterleaver.  While
this approach is obvious, it can be very slow compared to a purely hardware based solution and
it is by no means clear how the effective length of the
permutation can be reduced, without significant changes in the read/write order
of the elements and without significant degradation in the performance of the
resulting concatenated codes (in practice we need to satisfy both requirements). 

In \cite{fred1},
we have presented an alternative technique for the realization of the interleaver (deinterleaver)
as a finite state permuter (i.e., a serial queue implementing transpositions) whose operation was described via a transposition vector
that uniquely identifies the permutation. The representation of the permutation using
its transposition vector preserves information about the {\em prefix symbol substitution} property
of a permutation which directly correlates with how the interleaver (deinterleaver)
operates on the error events and their time shifted versions which are of
relevance in concatenated code design (see \cite{fred1} for
details).

We start this section by recalling some results from \cite{fred1}.
Consider an indexed set of elements $x_1 x_2 x_3 ...x_N$.
A given interleaver performs a particular permutation of this set of elements.
The permutation $\pi$ acts on the indices of the elements.
Henceforth, the notation $\pi (i) =j$ is used to mean that
the $j$-th input symbol is carried to the $i$-th position at the output.
It is a basic result in group theory \cite{hall} that any permutation
$\pi$ on a set of elements $S$ can be written as a product of
disjoint cycles, and $S$ may be
divided into {\em disjoint subsets} such that each cycle operates on
a different subset.
A cycle of length two is called a {\em transposition}. It is easy to verify
that any finite cycle can be written as a product of transpositions.
Hence, we conclude that transpositions represent the elementary
constituents of {\em any permutation}.

The Finite State Permuter (FSP) introduced in \cite{fred1},
is a realization of an interleaver in the form of a sliding window
transposition box of fixed length equal to the delay of the permutation
it implements on its input sequence. An FSP has the property
that a transposition
performed at a given time slot is responsible for the generation of the
output at the same time slot. Operation of the FSP
can be understood by thinking of
a sliding window as a queue. To generate any possible sequence of
outputs, it is sufficient to exchange the head of the queue with the element
that is to be ejected at that time slot.

\begin{figure}[t]
\centering
\includegraphics[width=3in]{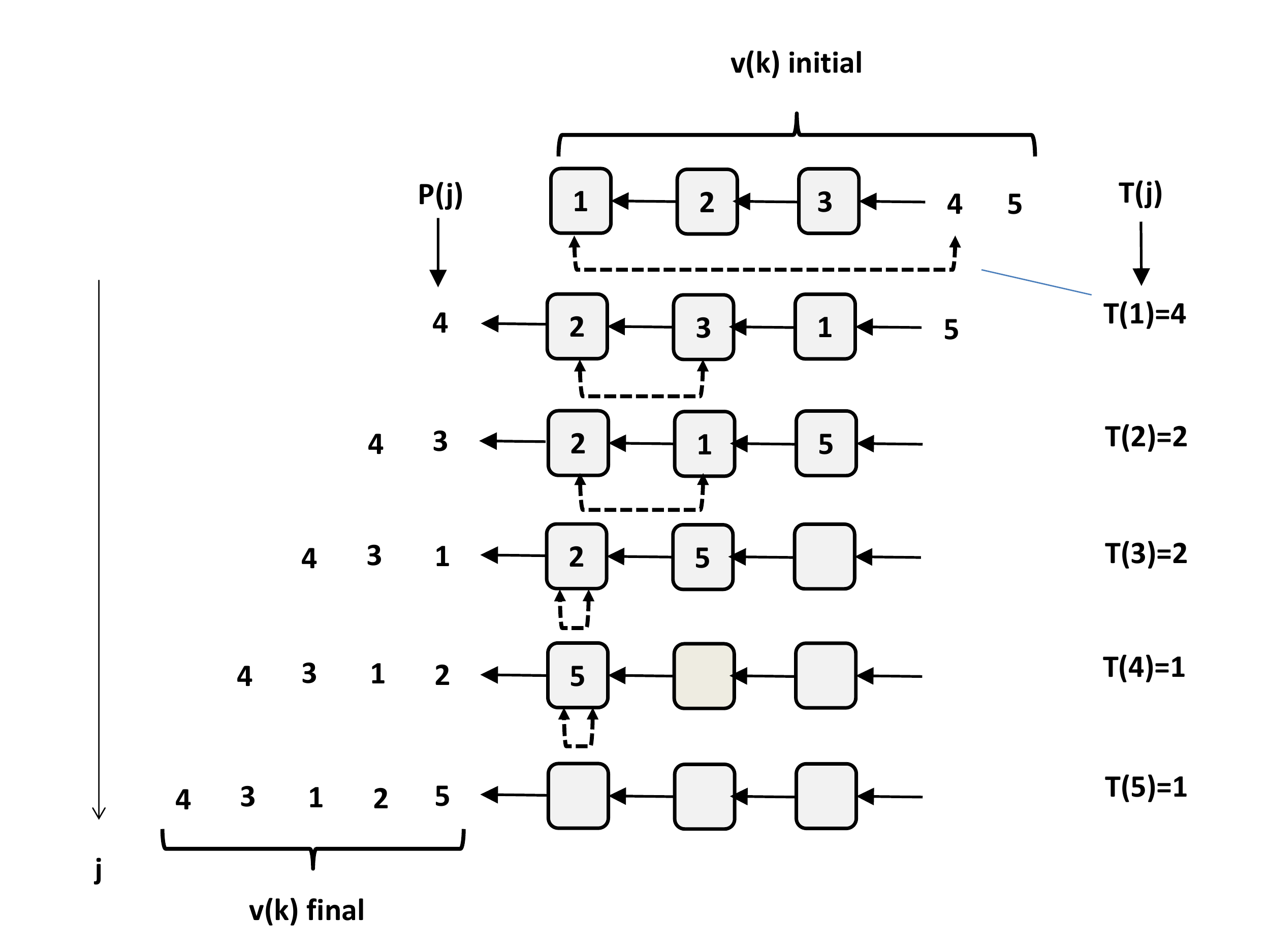}
\caption{Example of derivation of the transposition vector for a
given permutation. The memory cells are depicted as square boxes.
The sequence of transpositions is obtained by simply generating
the desired outputs from the input and the contents of the memory
cells one after another.} \label{ntras}
\end{figure}

Any permutation
on a finite set of elements can be represented using
a {\em unique transposition vector} associated with its FSP realization.
As an example consider the permutation
\begin{equation}
\pi=\left( \begin{array}{ccccc}
1 & 2 & 3 & 4 & 5 \\
4 & 3 & 1 & 2 & 5 \end{array} \right) .
\label{pi}
\end{equation}
Consider the queue model of an FSP and assume that data enters
from left to right as depicted in Fig.~\ref{ntras}.

Let us label the transpositions to be performed sequentially with
the head of the queue to generate the desired outputs, using
positive integers. Let the integer 1 denote the case whereby no
transposition is performed with the head of the queue and the
element at the head of the queue is simply ejected. Then the
transposition vector (to be read from left to right) that fully
defines permutation $\pi$ is $T_{\pi}=(4,2,2,1,1)$. The delay of
this permutation is 3 (i.e., one less than the largest element of the transposition vector), 
and the FSP can be implemented using three
memory cells. Any permutation on $N$ elements uniquely defines a
transposition vector of size $N$. Conversely, any transposition
vector of size $N$, defines a unique permutation on $N$ elements.
Note that when {\em synthesizing} a permutation using the
transposition vector $T$, the $k$-th element of vector $T$, when
scanned from right to left, can only assume values in the set
$\{1,2,...,k\}$.

The description of a permutation using the transposition vector of its
associated FSP realization turns out to be quite useful for
iterative interleaver construction and pruning.
This is because of a {\em prefix symbol
substitution} property that is best described through an example.
Consider the example transposition vector
$T_{\pi}=(4,2,2,1,1)$ associated with permutation $\pi$ above. Take
the binary sequence 10110 labeled from left to right.
Permutation $\pi$ maps this sequence
to 00111. Consider a new transposition vector
$T_{\pi_2}=(3,T_{\pi})=(3,4,2,2,1,1)$ associated with the permutation
\begin{equation}
\pi_2 =\left( \begin{array}{cccccc}
1 & 2 & 3 & 4 & 5 & 6 \\
3 & 5 & 4 & 2 & 1 & 6 \end{array}\right).
\label{pi2}
\end{equation}
Note that $\pi_2$ looks quite different from $\pi$, yet its output
corresponding to the {\em shifted} sequence $010110$ is
$001110$ which is a shifted version of the output generated by $\pi$.
In essence, the descriptions of $\pi$ and $\pi_2$ using the
transposition vectors preserves information about the
{\em prefix symbol substitution}
property of these two permutations on error patterns whereby the
first transposition exchanges a zero with a zero, or a one
with a one.

By its very nature a FSP can operate in continuous un-interrupted mode. Data simply
enters the FSP from right, there is an initial delay equal to the permutation delay after
which a sequence of transpositions are applied continuously. As the permuted sequence exits the FSP,
a new block of data to be permuted enters sequentially until the FPS buffer is full again and the
sequence of transpositions is repeated. Such a continuous flow mode is generally un-achievable
with any RAM structure which requires us to read the entire content of the RAM, wait until the RAM
is re-loaded and the process is repeated. Since pruning boils down to the truncation
of the transposition vector, this continuous flow property is maintained regardless of
the pruning length. In other words, the block length can change continuously without
incurring any delay in changing the interleaving length.  

\section{Basic Results on Interleaver Pruning and Growth}
We need to be able to go from one representation of a permutation to another easily. The first pseudo-code
below generates the transposition vector corresponding to a given permutation of length N. We start
with a vector of labels v=[1,2,3,...,N] and transform this via suitable transpositions to the permuted
set of labels as specified by the permutation $P$ (we call this routine, {\em perm2trans}):\\
\begin{verbatim}
for j=1 to N;
    for k=j to N;
    if v(k)=P(j), then
        T(j)=k-j+1;
        tmp=v(k);
        v(k)=v(j);
        v(j)=tmp;
        break
    end
    end
end
\end{verbatim}
The complexity of this routine is upper bounded by 0.5(N+1)N, i.e., it is quadratic in N. The main reason for this
is that at each step the contents of the buffer holding the data needs to be examined to find the next element
that needs to be swapped with the head of the queue to get the desired output. The next pseudo code generates
the permutation vector from the transposition vector. Here, we simply perform swapping of the elements in the queue
sequentially as dictated by the transposition vector to come up with the desired permutation (we call this routine, {\em trans2perm}):\\
\begin{verbatim}
for j=1 to N;
    s=P(j);
    P(j)=P(T(j)+j-1);
    P(T(j)+j-1)=s;
end
\end{verbatim}
Clearly the complexity of this routine is O(N).   

Next, suppose a permutation $\pi_N$ of length N with transposition vector $T_{N}$ is {\em grown} to a permutation of
length $(N+1)$, $\pi_{N+1}$ via growing the transposition vector to $T_{N+1}=[j,T_N]$, whereby $j\in \{1,2,..,N+1\}$. the
question is what is the relationship between $\pi_{N+1}$ and $\pi_N$? We have the following result in connection with this:\\
{\em Theorem-1}: 
\begin{eqnarray*}
\pi_{N+1}(1) = j & & \\
\mbox{if}\; j \neq 1,\; \mbox{then}  && \\
\pi_{N+1}(\pi_N^{-1}(k)+1) =k+1 && k=1,2,..,N\;\& \\
& & k\neq (j-1); \\
\pi_{N+1}(\pi_N^{-1}(j-1)+1)=1 && \\
\mbox{otherwise,}\\
\pi_{N+1}(\pi_N^{-1}(k)+1)=k+1, &&  k=1,2,..,N.
\end{eqnarray*}
{\em Proof}:  \\
The first transposition is $j$ and we swap the element at position 1 with $j$. The order of the elements in the {\em queue} (see figure~\ref{ntras}) becomes
$v=[2,3,...,((j-1),1,(j+1),...,N.N+1]$ where element 1 is in the $(j-1)$-th position from left in vector $v$ that is now of length $N$. The rest of the
transposition vector $T_N$ executes $\pi_N$ on the elements of $v$ and it is easy to see that:\\
element 2 appears at $(\pi_N^{-1}(1)+1)$-st position in the output sequence;\\
element 3 appears at $(\pi_N^{-1}(2)+1)$-st position in the output sequence, and so on; \\
element 1 appears at $(\pi_N^{-1}(j-1)+1)$-st position in the output sequence provided $j\neq 1$. If in fact $j=1$, we simply eject
element 1 and vector $v$ above is an ordered sequence from 2 to N and $(\pi_N^{-1}(k)+1)=k+1 \; \forall \; k=1,2,3,..,N$. $\Box$ \\
An example illustrates the process. Suppose $T_N=[4,2,2,1,1]$ and $T_{N+1}=[3,T_N]$. Using the {\em trans2perm} routine we obtain the
two permutations $\pi$ and $\pi_2$ in equations~(\ref{pi},\ref{pi2}) above. To verify the result on $\pi_2$ using the result of Theorem-1 above, 
we need the inverse permutation $\pi^{-1}$ which is given by:
\begin{equation}
\pi^{-1}=\left( \begin{array}{ccccc} 
1 & 2 & 3 & 4 & 5 \\
3 & 4 & 2 & 1 & 5 \end{array} \right) 
\end{equation}
We note that $j=3\neq 1$, hence we get:\\
\begin{eqnarray*}
\pi_2 (1) &=& 3 \\
\pi_2 (\pi^{-1}(1)+1)=\pi_2 (4) &=& 1+1=2 \\
\pi_2 (\pi^{-1}(2)+1)=\pi_2 (5) &=& 1 \; \mbox{since 2=j-1} \\
\pi_2 (\pi^{-1}(3)+1)=\pi_2 (3) &=& 3+1=4 \\
\pi_2 (\pi^{-1}(4)+1)=\pi_2 (2) &=& 4+1=5 \\
\pi_2 (\pi^{-1}(5)+1)=\pi_2 (6) &=& 5+1=6 
\end{eqnarray*}
The pruning process is just the opposite of the growth whereby we prune the interleaver of length $(N+1)$ with
transposition vector $T_{N+1}=[j,T_N]$ by dropping the first transposition $j$. From Theorem-1 the following result is immediate:\\
if $j\neq 1$, then:\\
\begin{eqnarray*}
\pi_N^{-1}(j-1) &=& \pi_{N+1}^{-1}(1)-1 \\
\pi_N^{-1}(k) &=& \pi_{N+1}^{-1}(k+1)-1, \; k=1,2,..,N \;\& \\
& &  \; \; \; \; \; \; \; \; \; \; \; \; \; \; \; \; \; \; \; \; \;  \; \; \; \; \; \; \; k\neq (j-1)
\end{eqnarray*}
if $j=1$, then:\\
\begin{eqnarray*}
\pi_N^{-1}(k) &=& \pi_{N+1}^{-1}(k+1)-1, \; k=1,2,..,N 
\end{eqnarray*}   
We can think of the transposition vector as an operator acting on a vector $s$ and producing a vector with 
permuted elements $\hat{s}$ (we could think of the
permutation really the same way). Notationally, we can write $\hat{s}=T_{N+1}(s)$ where, $s$ is a vector of elements of length $(N+1)$ 
and $s'$ is the permuted vector. Clearly, if $T_{N+1}^{-1}$ is the transposition vector of the inverse permutation, we must have
$T_{N+1}^{-1}T_{N+1}(s)=s$, hence, we can write $T_{N+1}^{-1}T_{N+1}=T_{I_{N+1}}$ where $T_{I_{N+1}}=[1,1,..1]$ is the transposition
vector of the identity permutation. Given how the transposition vector operates on the elements of vector $s$ in the queue, it is possible to derive the following result:
\begin{eqnarray}
\mbox{if}\; T_{N+1} &=& [j,T_1] \; \mbox{and} \\
T_{N+1}^{-1} &=& [k,T_2], \\
\mbox{then} \; T_1 T_2 &=& T_{I_{N}} \\
\mbox{consequently}, T_1=T_N & & T_2=T_N^{-1}
\label{transpart}
\end{eqnarray}   
in other words, $T_N$ and $T_N^{-1}$ are the transposition vectors of the pruned permutation and its inverse. 
%
The explanation for this results is as follows. The first transposition $j$ swaps element 1 with $j$. The content of the
queue after this transposition is $v=[2,3,...,(j-1),1,(j+1),...,N,N+1]$. To determine whether $T_1$ and $T_2$ are inverses
of each other, we relabel these elements into $\hat{v}=[1,2,...,(j-2),(j-1),j,...,N]$ and see the effect after the application of $T_1$ and 
$T_2$. The position of the element 1 in the queue after the application of $T_{N+1}$ must be $k$ since
$T_{N+1}^{-1}=[k,T_2]$. After the first swap the rest of the elements in the queue must be from 2 to $(N+1)$ in a scrambled order. 
Relabeling these elements again
by subtracting one from each element without changing their scrambled order we end up with the set of elements from 1 to N. 
Now, the combined action of $T_{N+1}$ and
$T_{N+1}^{-1}$ should restore the sequence to its original order. This is only possible if the relabeled sequences are permuted and
subsequently put back in successive order after the application of $T_1$ followed by $T_2$. This by definition implies that 
$T_1=T_N, \; T_2=T_N^{-1}$.   
 
The parsing of
the transposition vector can occur at any point desired and the result would continue to hold. In other words:
\begin{eqnarray*}
\mbox{if} \;\; T_{N} &=& [j_1,j_2,..j_{n1},T_1] \; \mbox{and} \\
T_{N}^{-1} &=& [k_1,k_2,..,k_{n1},T_2], \\
\mbox{then} \; T_1 T_2 &=& T_{I_{N-n1}} \\
\mbox{consequently}, T_1=T_{N-n1} & & T_2=T_{N-n1}^{-1}
\end{eqnarray*}   
In light of this and given our result on impact of pruning, we can now give a result on pruning in terms of the permutation itself
and not its inverse. In particular, if $T_{N+1}=[j,T_N] \; \mbox{and} \; T_{N+1}^{-1}=[k,T_N^{-1}]$, we can prove that:\\
{\em Theorem-2:}\\
if $k\neq 1$, then:\\
\begin{eqnarray*}
\pi_N (k-1) &=& \pi_{N+1}(1)-1 \\
\pi_N (n) &=& \pi_{N+1}(n+1)-1, \; n=1,2,..,N \;\& \\
& &  \; \; \; \; \; \; \; \; \; \; \; \; \; \; \; \;  \; \; \; \; \; \; \;  \; \; \; \; \; n\neq (k-1)
\end{eqnarray*}
if $k=1$, then:\\
\begin{equation}
\pi_N (n) = \pi_{N+1}(n+1)-1, \; n=1,2,..,N 
\label{casek1}
\end{equation}
{\em Proof:} is obvious in light of Theorem-1 and the relations~(\ref{transpart}). $\Box$
\subsection{Impact of Pruning on Permutation Spread}
One important issue concerns the spread of the permutation which appears to play a key role for instance in the performance of Turbo
codes. In this subsection, we wish to examine the impact of our pruning method on the spread properties of the underlying permutations.
Given the recursive nature of the pruning and growth processes, we anticipate to develop recursive relationships for the impact of pruning
and growth on permutation spread. First, let us give a generic expression for the permutation spread. Let $T_{N+1}=[j,T_N] \; \mbox{and} \; T_{N+1}^{-1}=[k,T_N^{-1}]$ be the transposition vectors of permutation $\pi_{N+1}$ and its inverse. We define the spread of permutation
$\pi_{N+1}$, denoted $sprd(\pi_{N+1})$, as follows:
\begin{equation}
sprd(\pi_{N+1})=\min_{n_1,n_2}|\pi_{N+1}(n_2)-\pi_{N+1}(n_1)|+|n_2 -n_1|
\end{equation} 
Given our general result above on pruning, we have the following result on spread properties of pruned interleaver: \\
{\em Theorem-3}: \\
if $k\neq 1$, let:
\begin{equation}
\alpha=\min_{m=1,2,..,N,\; m\neq (k-1)} [|\pi_{N+1}(m+1)-\pi_{N+1}(1)|+|m -k+1|]
\label{alph}
\end{equation}
then,
\begin{eqnarray*}
sprd(\pi_N) &\geq & sprd(\pi_{N+1})\; \mbox{if} \;\; \alpha \geq sprd(\pi_{N+1}) \\
sprd(\pi_N) &=& \alpha \; \mbox{if} \;\; \alpha < sprd(\pi_{N+1})
\end{eqnarray*}
if $k=1$, then:\\
\begin{equation}
sprd(\pi_N) \geq sprd(\pi_{N+1})
\end{equation}
{\em Proof:} \\
Note that if $k\neq 1$:
\begin{eqnarray*}
|\pi_{N}(n_2)-\pi_{N}(n_1)|+|n_2 -n_1| &=& \\
|\pi_{N+1}(n_2+1)-\pi_{N+1}(n_1+1)|+|(n_2+1) -(n_1+1)| & & \\
\forall \; n_2,n_1=1,2,..N,\; n_2,n_1\neq (k-1) & &
\end{eqnarray*}
minimization of $|\pi_{N}(n_2)-\pi_{N}(n_1)|+|n_2 -n_1|$ over all $n_1,n_2$ except for $n_1 \;\mbox{or}\; n_2=(k-1)$ is the same as minimization
of $|\pi_{N+1}(n_2+1)-\pi_{N+1}(n_1+1)|+|(n_2+1) -(n_1+1)|$ over the same range of indexes. Since the search space is reduced relative to the
case where we search for minimum of $|\pi_{N+1}(n_2)-\pi_{N+1}(n_1)|+|n_2 -n_1|$ over all possible values of $n_1,n_2=1,2,..,(N+1)$, the result
can only be a number greater than or equal to $sprd(\pi_{N+1})$. The only remaining case that needs to be examined, is when either $n_1=(k-1)$ or $n_2=(k-1)$ and
that leads to our definition of $\alpha$.\\
If $k=1$, then from equation~(\ref{casek1}), the set of numbers used to find the spread is really the same for $\pi_N$ and $\pi_{N+1}$ except that
the search space for finding the spread of $\pi_{N+1}$ is larger, hence the computed minimum for $\pi_{N+1}$ can only be equal or smaller and therefore the spread of $\pi_{N}$ can only be equal or larger than that of $\pi_{N+1}$. $\Box$\\
\subsection{Impact of Pruning on Algebraically Constructed Interleavers}
A variety of algebraically constructed permutations have been proposed in the literature. The idea is to have a formula
to construct permutations having certain desirable properties (e.g., high spread). The main requirement in any algebraic mapping to produce a permutation is closure and the requirement that no two integers are mapped via the permutation to the same number (i.e., the function is bijective).
In connection with such permutations, the main result of interest is Theorem-2 and in particular, the relation $\pi_N (n)=\pi_{N+1}(n+1)-1$ which
is a linear relationship suggesting that the core property that the mapping is obtained through an algebraic function, for
the most part, remains intact after pruning. 

An example class of algebraic permutations are the Quadratic Permutation Polynomials (QPPs) \cite{take1,take2} of length $K$ defined through the relationship:
\begin{equation}
\pi_K(j)=j \cdot h+j^2 \cdot b+c \pmod K  \; ,j=0,1,...,(K-1).
\end{equation}
Note that in this paper, the permutation starts with index 1, while in above definition, the first index is 0. To be consistent with
our indexing, we can easily write:
\begin{equation}
\pi_K(j)=(j-1) \cdot h+(j-1)^2  \cdot b+c \pmod K +1 \; ,j=1,...,K.
\end{equation}
where, there are two cases to consider for the construction to be valid. In case-1, $\gcd(h,K)=1$ and let $K=\prod_i p_i^{n_i}$ be the prime factorization of $K$. If 2 is not a prime factor or a prime factor with power greater than or equal to 2, then the prime factors of {\em b} 
must include at least all prime factors of $K$ other than 2 (possibly with different powers), 
$b=m\prod_i p_i^{m_i}, \; m \nmid  p_i \; \& \; m_i\geq1$. In case-2, 2 is a prime factor with power of 1, $\gcd(h,K/2)=1$, and $h+b$ must be
odd, $b=m\prod_i p_i^{m_i}, \; m \nmid  p_i \; p_i\neq 2, \& \; m_i\geq1 \; \& \; 2\nmid m$ if $h$ is even. A toy example is
$K=3\times 5$, $h=2, \; \gcd(h,K)=1, \; b=3\times 5$ leading to $\pi_{15}(j)=2j+15j^2 +c\; \pmod{15},\; j=0,1,2,..,14$, where, $c<K$
is an arbitrary shift parameter so that $\pi_{15}(0)=c$. 

\begin{figure}[t]
\centering
\includegraphics[width=3in]{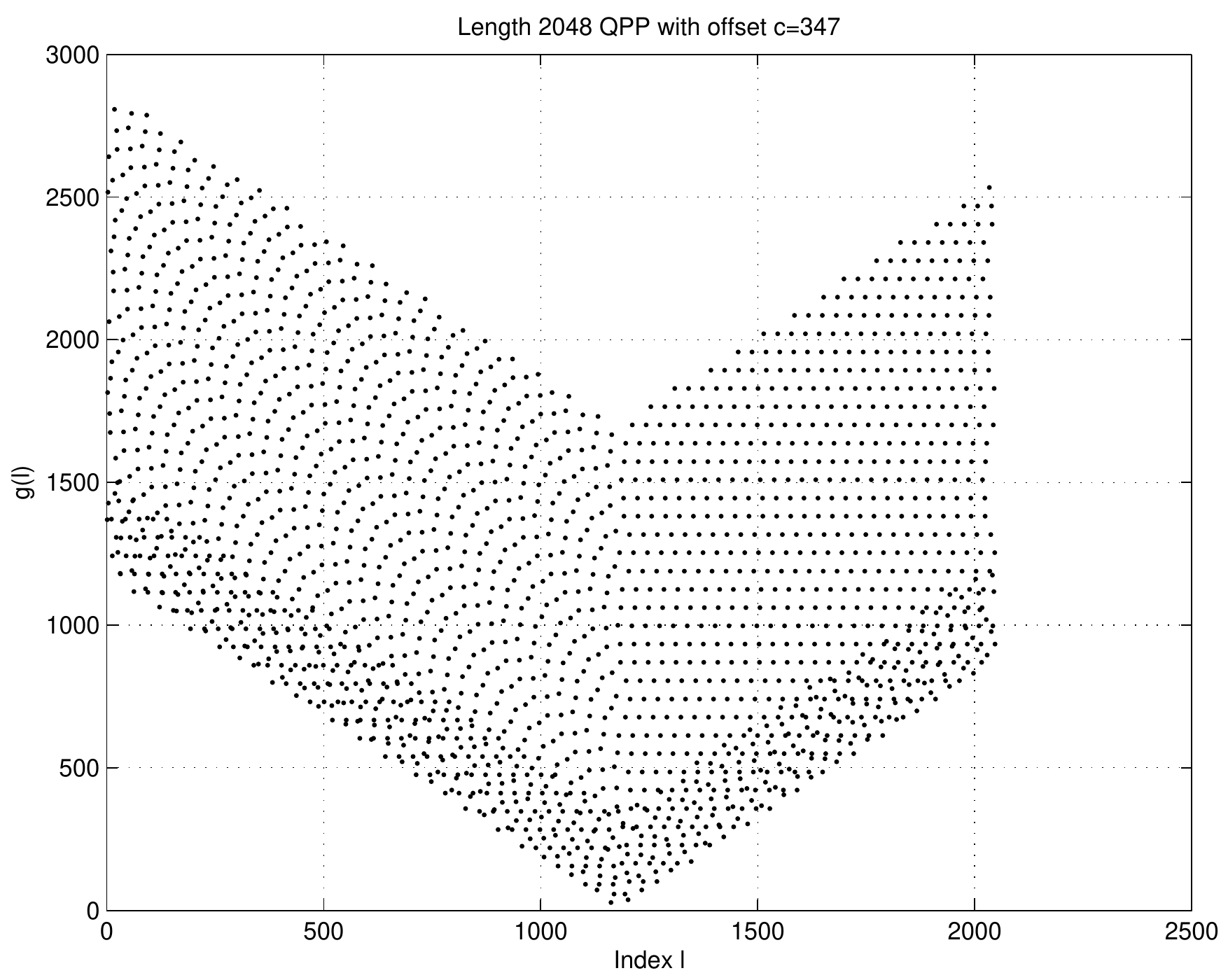}
\caption{Scatter plot of $g(l)$ for a length 2048 QPP permutation with offset $c=347$.} \label{qpp1}
\end{figure}

\begin{figure}[t]
\centering
\includegraphics[width=3.5in]{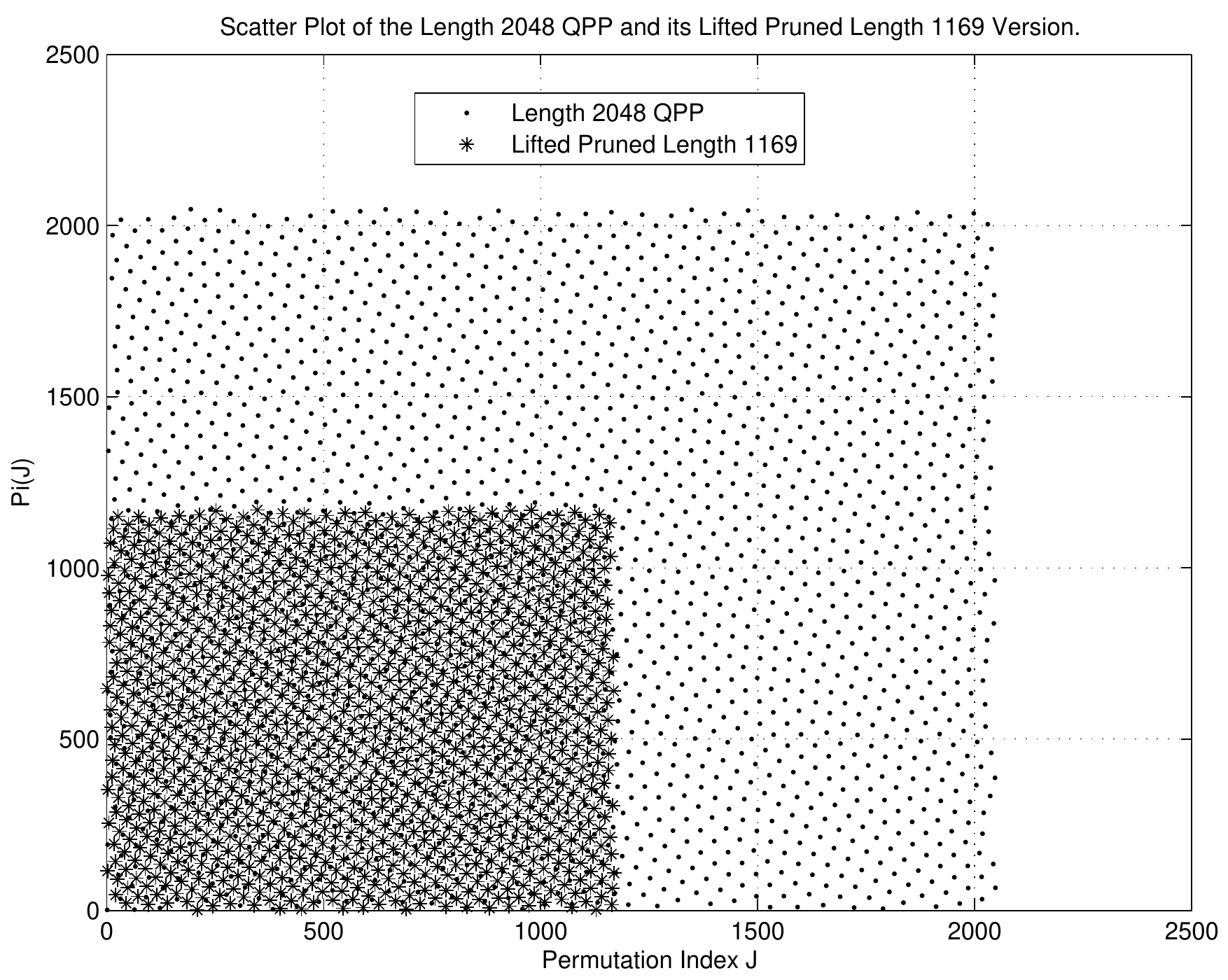}
\caption{Scatter plot of the length 2048 QPP mother permutation with offset $c=0$ and the
lifted pruned permutation with $M=500$ and effective length 1169.} \label{qppfig1}
\end{figure}

\begin{figure}[t]
\centering
\includegraphics[width=3.5in]{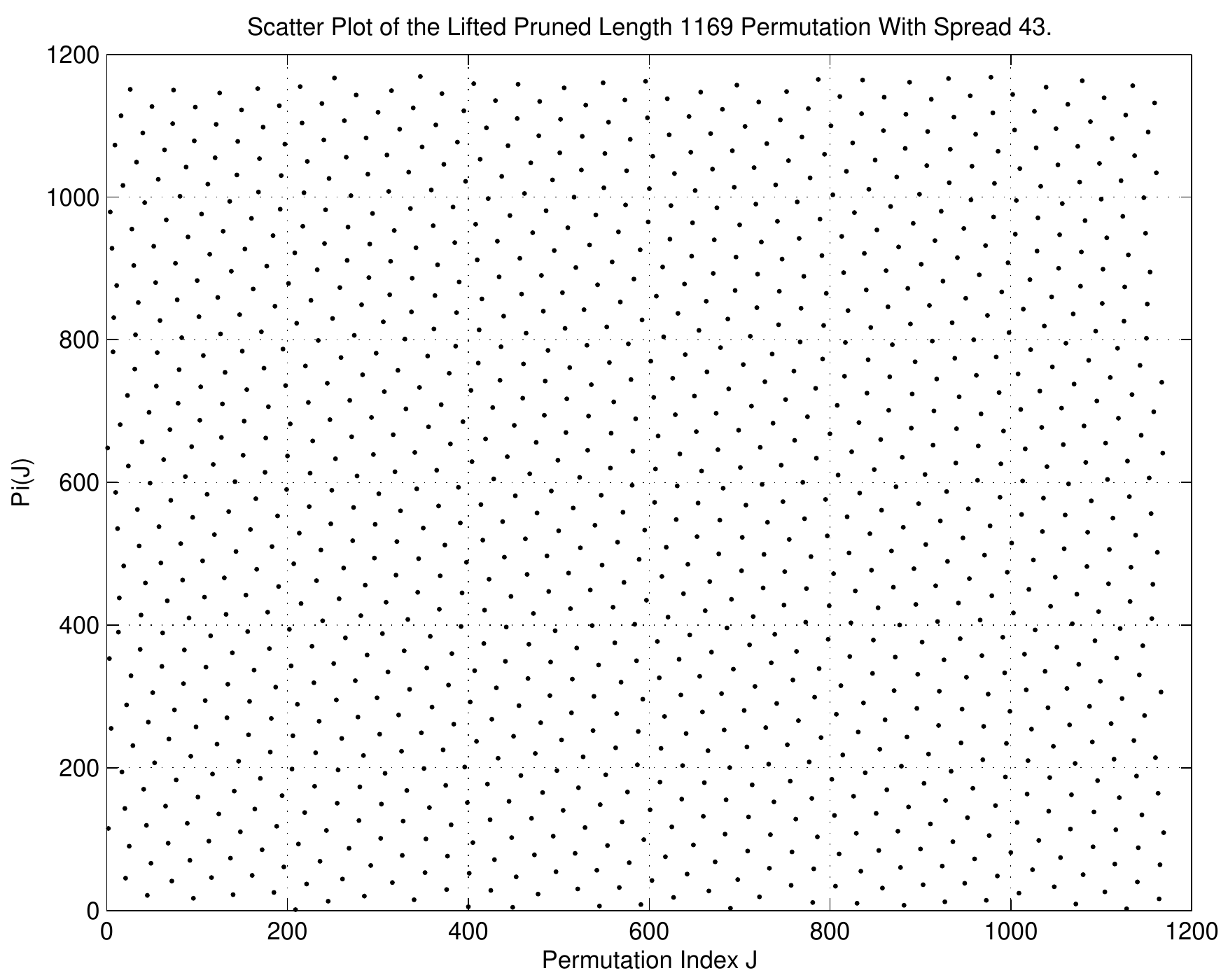}
\caption{Scatter plot of the length 1169 lifted pruned permutation. The QPP mother permutation has length 2048 and offset $c=0$.} 
\label{qppfig2}
\end{figure}

The transposition vector of a generic QPP is easily seen to be of the form $T_{K}=[c+1,T_{K-1}]$. Let the transposition vector of the inverse
permutation be $T_{K}^{-1}=[n,T_{K-1}^{-1}]$, then:\\
if $n\neq 1$:
\begin{eqnarray*}
\pi_{K-1}(n-1) &=& \pi_K (1)-1=c \\
\pi_{K-1}(l) &=& \pi_K (l+1)-1=l \cdot h+l^2  \cdot b+c \pmod K ,\\
 & & l=1,2,...,(K-1), \; l\neq(n-1).
\end{eqnarray*}
if $n=1$:
\begin{eqnarray*}
\pi_{K-1}(l) &=& \pi_K (l+1)-1=l \cdot h+l^2  \cdot b+c \pmod K ,\\
 & & l=1,2,...,(K-1).
\end{eqnarray*}
The algebraic structure of the permutation is indeed retained, albeit, with the same modulus operation as before. It is not difficult to show
that the repeated iteration of the pruning via truncation of the transposition vector always leads to a permutation that
is algebraically defined via the same equation but with shifted indexes over a smaller subset of overall indexes (i.e., that
is the only way the permutation can remain to be a valid bijective map). Indeed, suppose a QPP permutation of length $K$ is
pruned by truncating its transposition vector by $M$ positions. Repeated application of the above result suggests that
there is a large fraction of the permutations (provided $M/K$ is small) that must satisfy the following:
\begin{equation}
\pi_{K-M}(l)=(l+M-1) \cdot h+(l+M-1)^2 \cdot b+c \pmod K +M-1
\label{qppt}
\end{equation} 
In terms of the
spread properties of the pruned interleaver, the key parameter to compute is $\alpha$ as defined in equation~(\ref{alph}):\\
if $n\neq 1$:
\begin{eqnarray*}
\alpha &=& \min_{S} [|\pi_{K}(l+1)-\pi_{K}(1)|+|l -n+1|] \\
 &=& \min_{S} [|l \cdot h+l^2  \cdot b+c \pmod K -c|+|l -n+1|] \\
\end{eqnarray*}
assuming $c<(K-1)$, where the set $S=\{l=1,2,..,(K-1),\; l\neq (n-1)\}$. For a fixed set of parameters $(h,b)$, we can easily obtain
a plot of $g(l)=|l \cdot h+l^2 \cdot b+c \pmod K -c|+|l -n+1|$, then for any given $n$, $\alpha$ is the minimum of $g(l)$. As an example, take the QPP
$\pi(j)=63j+128j^2+347 \pmod{2048}$, then $g(l)=|63l+128l^2 +347 \pmod{2048} -347|+|l -1180+1|$ is plotted in figure~\ref{qpp1}. 
We note that
repeated application of the pruning based on truncation of the transposition vector can have a catastrophic impact on the spread
properties of the pruned interleaver, i.e., maintenance of the algebraic property of the permutation is no guarantee of its continued
performance when it comes to having high spread. However, in light of equation~(\ref{qppt}), this problem can be remedied as follows:\\
\begin{enumerate}
\item Much of the spread properties of the QPP permutations (and indeed many other algebraically constructed permutations), comes from
the good distribution of points in the XY plane when we plot points with coordinates $(j, \pi(j))$. Such properties are embodied
in the original algebraic expression used in definition of the permutation itself;
\item Truncation of the transposition vector of the permutation
as proposed here for construction of variable length interleavers may cause folding of certain number of points due to
the equation $\pi_{K-1}(n-1)=\pi_K (1)-1=c$ when $n\neq 1$, in the same region containing the points satisfying the core algebraic relationship
defining the mother permutation. The effect on spread can be catastrophic because of such folding. To illustrate, consider the QPP
mother permutation of length 2048 defined via:
\begin{eqnarray*}
\pi(i)&=&63\cdot(i-1)+128\cdot(i-1)^2+c\pmod{2048}+1 \\
 i &=& 1,2,...,2048.
\end{eqnarray*}
and suppose we truncate the transposition vector of this permutation by just 10 positions. The scatter diagram of the resulting permutation of 
length 2038 is shown in figure~\ref{qppcat1}.\\
\begin{figure}[t]
\centering
\includegraphics[width=3.5in]{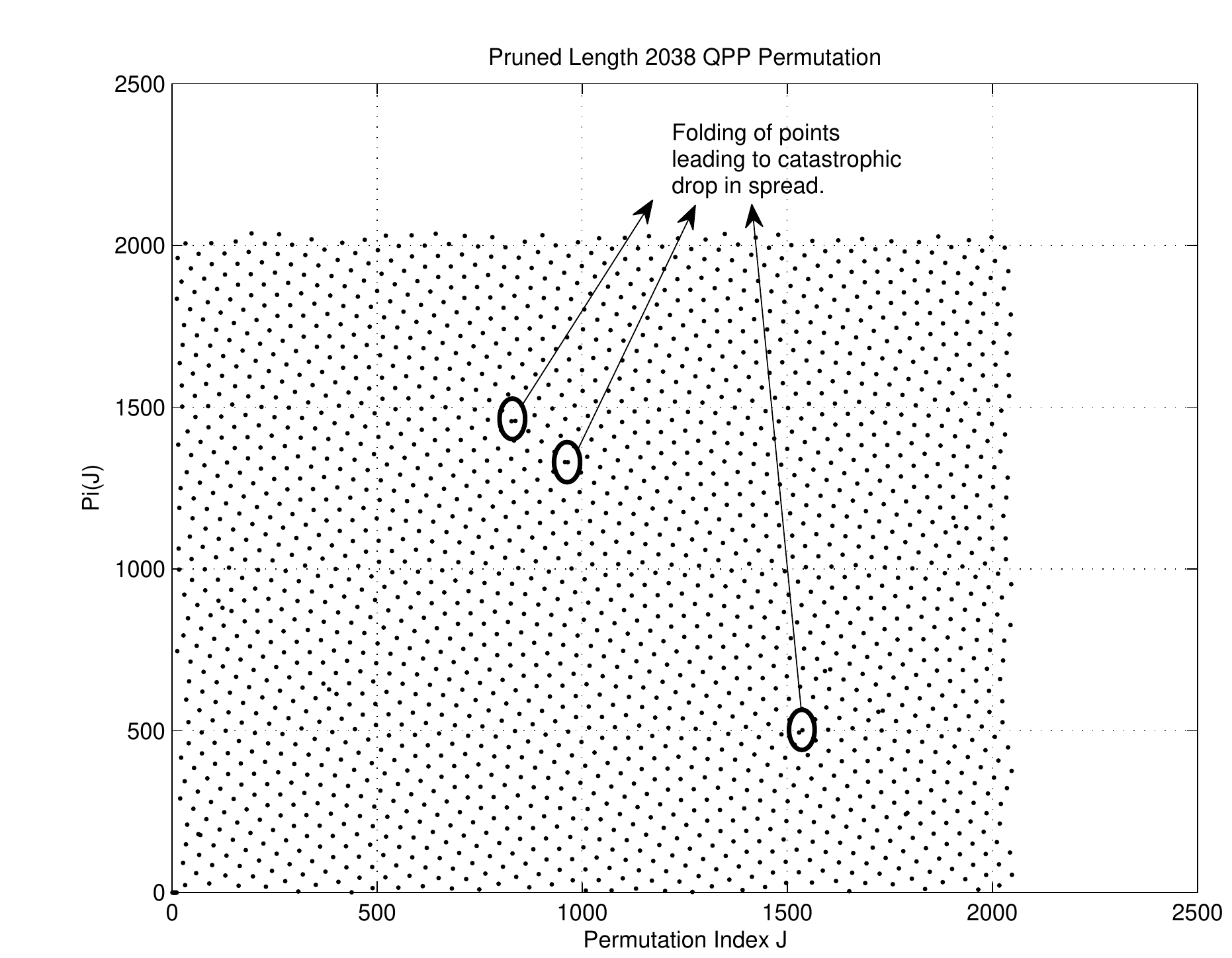}
\caption{Scatter plot of the length 2038 pruned QPP permutation with offset $c=0$.} \label{qppcat1}
\end{figure}
The folding of the points causes the spread of the resulting permutation to drop from 64 for the length 2048 mother permutation down to 2, 
even though almost all the other 
points are well distributed in the XY plane, hence, leading to high spreads;
\item The folding problem can be easily resolved in effect by {\em lifting} such points hence reducing the permutation size from its
original target length, but restoring the spread properties of the pruned permutation. There are two issues to address, a)~ how do we identify
such points?, and b)~how do we actually implement a systematic variable length interleaver using the transposition vector paradigm? The
identification of the valid points from those we would like to lift is actually quite straightforward, we simply check and
see if the relation $\pi_{K-M}(l)=(l+M-1)\cdot h+(l+M-1)^2 \cdot b+c \pmod K +M-1$ holds for index $l$, if it does not, we flag the index as invalid and
lift the point. The question of how do we actually lift the point in practice in the context of sequential operation of the finite
state permuter is trickier. To handle this problem we proceed as follows:\\
\begin{itemize}
\item in the sequential entry of data in the queue used to hold the data and implement the transpositions we insert dummy symbols at
locations where we wish to lift the points from the permutation map;
\item we truncate the transposition vector to an initial target length $(K-M)$ by eliminating the first $M$ transpositions from left,
in the transposition vector of the mother interleaver;
\item we apply the truncated transpositions to the data in queue containing the dummy as well as valid symbols to be permuted and run the
FSP to generate the output. We then simply discard the dummy symbols from the output sequence and get the permuted data exhibiting 
the desirable spread properties.
\end{itemize} 
\end{enumerate} 
An example illustrates the procedure for interleaver pruning with lifting that leads to maintenance of the desirable spread properties
of the pruned interleaver. Take the QPP defined via:
\begin{eqnarray*}
\pi(i) &=& 63.(i-1)+128.(i-1)^2 +c\pmod{2048} +1 \\
 i &=& 1,2,...,2048.
\end{eqnarray*}
and let the offset $c=0$ for convenience (this does not change the spread properties of the pruned permutations). Suppose we set $M=500$ and
truncate the transposition vector of the above permutation by 500 positions. The scatter plot of the mother length 2048 QPP permutation and
the lifted pruned permutation is shown in figure~\ref{qppfig1}.
While the target length of the pruned interleaver is $(2048-500)=1548$, the actual length of the pruned interleaver after lifting
is 1169 and hence, 379 points in the permutation map had to be lifted. The original spread of the mother permutation is 64, while that
of lifted pruned interleaver is 43. If lifting was not done, the spread of the length 1548 permutation would have been 2. The scatter plot
of the length 1169 lifted pruned permutation showing regular and well disbursed spread of points in the XY plane is shown in figure~\ref{qppfig2}.
\section{Conclusions}
Systematic methods of constructing variable length interleavers with application to variable-rate, delay and decoding complexity
channel coding and possibly other application areas, are rare. On the surface, the problem may be non-existent in that 
one could hypothetically design a set of interleavers with varying lengths and store the information and use a purely software
approach by writing and reading into a RAM. There are several drawbacks to such an approach, one is in terms of implementation 
complexity that can be significantly higher than more hardware oriented approaches, the other is speed whereby
a hardware based solution can generally run much faster than a RAM based approach and finally the fact that 
one must design the system a-priori and by its very nature the design is inflexible
(i.e., one could not change the interleaving block length at will). The idea of using a good mother interleaver and 
constructing a variable length interleaver out of it via a clever strategy is clearly appealing and that is what this paper
and the approach presented in \cite{fredprune, mans1, mans2, mans3} is all about. In coding, a good low rate code is 
punctured to get higher rate codes.
In this paper, the transposition vector of a good interleaver can be truncated to get lower block length and hence delay
interleavers. Of course, one must define what the goodness criterion is. In this paper we have focused on permutation spread
which broadly speaking plays a key role in concatenated coding applications.

\vspace{1cm}

{\bf Fred Daneshgaran} received the B.S.  degree in electrical and mechanical engineering 
from California State University, Los Angeles (CSLA) in 1984, the M.S. degree in electrical 
engineering from CSLA in 1985, and the Ph.D. degree in electrical engineering from 
University of California, Los Angeles (UCLA), in 1992.

From 1985 to 1987, he was an Instructor with the Department of Electrical and Computer 
Engineering (ECE) at CSLA. From 1987 to 1993, he was an Assistant Professor, 
from 1993 to 1996, he was an Associate Professor, and since 1997, he has been a 
full Professor with the ECE Department at CSLA. Since 1989, he has been the 
Chairman of the Communications Group of the ECE Department at CSLA. 
From 2006 he serves as the chairman of the ECE department.

Professionally, from 1999 to 2001, he acted as the Chief Scientist for TechnoConcepts, Inc., 
where he directed the development of a prototype software-defined radio system, managed the 
hardware and software teams, and orchestrated the entire development process. 
In 2000, he co-founded EuroConcepts s.r.l., an R\&D company specializing in the design of 
advanced communication links and software radio. In 1996, he founded Quantum Bit 
Communications, LLC, a consulting firm specializing in wireless communications. 
Dr. Daneshgaran is the director of the fiber and non-linear optics research 
laboratory at CSLA. The laboratory development was funded by the National Science 
Foundation (NSF) and the Office of Naval Research (ONR). This  
facility is focused on conducting R\&D in the general areas of sensory applications 
of fiber optics and optical information processing.

Dr. Daneshgaran served as the Associate Editor of the IEEE Trans. On Wireless Comm. 
in the areas of modulation and coding, multirate and multicarrier communications, 
broadband wireless communications, and software radio, from 2003 to 2009.
He has served as a member of the Technical Program Committee (TPC) on numerous 
conferences. Most recent contributions include IEEE WCNC 2013, CONWIRE 2012, ISCC 2012, 
ISCC 2011, and PIMRC 2011.

\vspace{1cm}

{\bf Marina Mondin} is Associate Professor at Dipartimento di Elettronica, Politecnico di Torino. 
Her current interests are in the area of signal processing for communications, modulation and coding, 
simulation of communication systems, and quantum communication. She holds two patents. 
She is Associate Editor for IEEE Transactions on Circuits and Systems-I. She has been acting as a 
reviewer for several international scientific IEEE and IEE journals, she has been Guest Editor for 
the EURASIP Journal on Wireless Communications and Networking (2009) and the International Journal 
of Digital Multimedia Broadcasting (2009 and 2010), and she has been member of the technical-scientific 
committees of various international conferences, such as the SNeS 2012, IADIS Collaborative 
Technologies conference 2010, 2011 and 2012, ICT 2010, ISWCS 2010, Spacomm 2009,2010 and 2011, 
Mobilight 2010, ICT 2009, Tridentcom 2008, Globecom 2006, Eusipco 2006, WHAPS 2005, WPMC 2004. 
She has organized an invited session on QKD at the conference ISABEL 2010 and ISABEL 2011. 
She has been in the 2012 TCAS committee for the selection of the Darlington and Guillemin-Cauer 
Best Paper Awards. 

She has been principal investigator for Politecnico di Torino in several national PRIN projects, 
funded by the Italian Ministry of Research and University, concerning the integration of satellites 
and high altitude platforms for broadband data transmission, she has been the National 
Coordinator of the PRIN 2007 project “Feasibility study of a Earth-satellite quantum 
optical communication channel”, and Co-PI for the NATO Collaborative Linkage Grant 
“Quantum photonics for secure quantum communication”, involving INRIM, Politecnico di Torino, 
Moscow Univ. and California State Univ., Los Angeles.

Prof. Mondin has authored and co-authored more than 50 articles on international journals and 
more than 100 contributions to international conferences. She has also cohauthored the books 
``Esercizi risolti di Comunicazioni Elettriche,'' CLUT (Cooperativa Libraria Universitaria Torinese), 
Torino, March 1997 and "Elaborazione Numerica dei Segnali", Pearson, 2007 (both in Italian).

\end{document}